\documentclass[12pt,letterpaper]{article}

\usepackage{amsmath,amssymb,calc}
\usepackage{graphicx}

\newcommand\be{\begin{equation}}
\newcommand\ee{\end{equation}}
\newcommand{\bea}{\begin{eqnarray}}
\newcommand{\eea}{\end{eqnarray}}

\newcommand{\nn}{\nonumber}
\newcommand{\pd}{\partial}

\def\id{\protect{{1 \kern-.28em {\rm l}}}}

\def\id{\protect{{1 \kern-.28em {\rm l}}}}

\begin{document}

\begin{titlepage}
\begin{center}
\hfill \\
\vspace{2cm}
{\Large {\bf Electroweak Symmetry Breaking from Gauge/Gravity Duality \\[3mm] }}

\vskip 10mm

{\bf Lilia Anguelova}

\vskip 4mm
{\em Department of Physics}\\
{\em University of Cincinnati, Cincinnati OH 45221, USA}\\
{\tt anguella@ucmail.uc.edu}

\vskip 6mm

\end{center}

\vskip .1in
\vspace{1cm}

\begin{center} {\bf Abstract}\end{center}

\vspace{-1cm}

\begin{quotation}\noindent

We use the gauge/gravity duality to study a model of walking technicolor. The latter is a phenomenologically promising framework for dynamical electroweak symmetry breaking. A traditional problem for technicolor models has been the need to address gauge theories at strong coupling. Recent developments in gauge/gravity duality provide a powerful tool for handling this problem. First, we revisit previously considered holographic models of QCD-like technicolor from D-branes. In particular, we develop analytical understanding of earlier numerical computations of the Peskin-Takeuchi S-parameter. Then we apply this method to the investigation of a model of walking technicolor, obtained by embedding ${\rm D}7$-$\overline{{\rm D}7}$ probe branes in a recently discovered type IIB background dual to walking behaviour. As a necessary step, we also show that there is an embedding of the techniflavor branes, that realizes chiral symmetry breaking. Finally, we show that the divergences that appear in the S-parameter can be removed by using holographic renormalization.

\end{quotation}
\vfill

\end{titlepage}

\eject

\tableofcontents

\section{Introduction}

Understanding electroweak symmetry breaking (EWSB), or equivalently the origin of mass, is a great challenge for phenomenology. In the Standard Model, EWSB is achieved via a fundamental scalar, the Higgs boson. Although conceptually simple, this possibility leads to well-known problems. More precisely, the masses of scalar fields are destabilized by quantum corrections and, even when stabilization is achieved via supersymmetry, there is an unnaturally large hierarchy between the electroweak and Planck scales. An appealing alternative to the Higgs boson is provided by the possibility of dynamical chiral symmetry breaking \cite{Tech}; models that explore that idea are known as technicolor models. 

The original technicolor proposals were simply scaled-up versions of QCD and so were incompatible with electroweak precision measurements \cite{PT}. However, later models \cite{WalkTech}, called walking technicolor as the relevant gauge couplings run slower than in QCD, are considered phenomenologically viable; for a pedagogical recent review, see \cite{MP}.\footnote{We should note the important role of \cite{AS} in the recent renewal of interest in walking technicolor.} A common problem, though, for all kinds of technicolor models is that the relevant physics occurs at strong coupling. Therefore, a direct field theory computation of various quantities of interest is not possible and so the present experimental bounds are insufficient to distinguish between a Higgs boson and a walking technicolor sector. Here we will address this problem by using a powerful recently-developed tool for studying gauge theories at strong coupling. Namely, we will utilize the gauge/gravity duality to study a model of walking technicolor.

Holographic models of regular technicolor\footnote{By 'regular' we mean the original QCD-like version, and not walking technicolor.} have been considered in \cite{HolTech}. These authors use D-brane configurations, very similar to the one giving the holographic QCD model of Sakai and Sugimoto \cite{SS,SS2}, in order to realize chiral symmetry breaking.\footnote{We should note that there is a large amount of work on a class of holographic technicolor models (loosely) inspired by AdS/CFT \cite{PhenoHolTech}, which however cannot be consistently embedded in string theory.} This is then translated to electroweak symmetry breaking via an appropriate embedding of the electroweak $SU(2)\times U(1)$ group into the techniflavor one. In this class of models one can compute the Peskin-Takeuchi S-parameter \cite{PT}, that is an important electroweak observable. In \cite{HolTech}, this computation was performed numerically. Before turning to walking technicolor, we will first revisit the above regular technicolor considerations with a more analytical approach. This will enable us to gain a better understanding of the situation. And also, it will be a useful preparation for the more involved new case.

To obtain a gravity dual of walking technicolor in the vein of \cite{HolTech}, we need, first, a gravitational background that is dual to a walking gauge theory and, second, a U-shaped D-brane embedding as in \cite{SS}, in order to achieve geometrical realization of chiral symmetry breaking. Fortunately, it was shown recently \cite{NPP} that a suitable background is provided by one of the type IIB solutions found in \cite{CNP}. The latter are deformations of the original Maldacena-Nunez background \cite{MN}, which still arise from D5 branes wrapping an $S^2$. In this gravity dual of walking behaviour \cite{NPP}, we will consider ${\rm D}7$-$\overline{{\rm D}7}$ probes and show that there is an embedding of a U-shape type a la Sakai-Sugimoto. Using this set-up as our model of walking technicolor, we will then compute the S-parameter with the methods we developed for the regular technicolor case.  

In the walking case, it will turn out that the answer for the S-parameter needs to be renormalized. This is, perhaps, not surprising since the Maldacena-Nunez background has long been known to lead to divergences. The novelty, however, is that, due to recent advances \cite{MM} in the program of holographic renormalization \cite{HolRen}, the background of interest for us can be renormalized. More importantly, we will renormalize the probe ${\rm D}7$ brane action, that we need, by adding an appropriate counterterm. This will then enable us to extract a finite answer for the renormalized S-parameter. Our analytical result gives us interesting insights. However, a numerical prediction for the value of S is hindered by the presence of a set of constants, that can only be fixed by numerical methods. The determination of those constants is work in progress \cite{ASW}.

The present paper is organized as follows. In Section \ref{HolTechSec}, we give a brief overview of the holographic construction of technicolor models from D-brane configurations in string theory. Furthermore, we review and slightly revise the computation of the S-parameter in this class of models. In Section \ref{RegTechEx}, we consider a model of regular technicolor, obtained by placing ${\rm D}7$-$\overline{{\rm D}7}$ probes in the conifold. We develop an analytical approach to solving the equations that determine the S-parameter, which gives us better insight into the numerical results of \cite{HolTech}. In Section \ref{WalkT}, we study a walking technicolor model obtained by embedding ${\rm D}7$-$\overline{{\rm D}7}$ probes in the background of \cite{NPP}. We use our new approach to extract an analytical answer for S. However, the latter turns out to contain IR divergences, that are the gravity dual of field theory UV divergences. So we add an appropriate counterterm and compute the renormalized S-parameter. Finally, in Section \ref{Disc} we discuss open issues and future research directions.  

\section{Technicolor from holography} \label{HolTechSec}
\setcounter{equation}{0}

The basic idea is the following. Let us consider a type II gravity background created by a certain number $N_{TC}$ of ${\rm D}q$-branes and embed in it $N_{TF}$ probe ${\rm D}p$-$\overline{{\rm D}p}$ branes. If there is an embedding of the probe branes of a U-shape form, like in the Sakai-Sugimoto holographic QCD model \cite{SS}, then one has a geometric realization of chiral symmetry breaking. Namely, the techniflavor group $U_L(N_{TF}) \times U_R(N_{TF})$, corresponding to separate ${\rm D}p$ and $\overline{{\rm D}p}$ branes, is broken to the diagonal subgroup $U(N_{TF})$ because of the joining of the ${\rm D}p$ and $\overline{{\rm D}p}$ at a certain position along the radial direction. Now, upon an appropriate embedding of the electroweak $SU(2)\times U(1)$ into $U_L(N_{TF}) \times U_R(N_{TF})$, the above chiral symmetry breaking translates into electroweak symmetry breaking. In the first two references of \cite{HolTech}, the technicolor branes were ${\rm D}4$-branes and the techniflavor probes were ${\rm D}8$-$\overline{{\rm D}8}$ as in \cite{SS}. The last reference in \cite{HolTech} considered several set-ups in both type IIA and type IIB.\footnote{Although such models, being dual descriptions of regular technicolor, naturally lead to unrealistically large values of the S-parameter, see \cite{AB} for recent progress in the direction of reducing somewhat the S-parameter value within the context of the Sakai-Sigimoto-like D4/D8 brane set-up.} To obtain a model of walking technicolor, we will embed ${\rm D}7$-$\overline{{\rm D}7}$ probes in the background of \cite{NPP}. 

Since our main concern will be the S-parameter, let us now recall its holographic computation. We will mostly follow the last reference in \cite{HolTech}, whose general treatment encompasses the rest of the references there. However, we will be a bit more careful than \cite{HolTech} and so will obtain a slightly different formula. 

\subsection{S-parameter: generalities}

The Peskin-Takeuchi S-parameter \cite{PT} is defined as:
\be \label{Sdef}
S = - 4 \pi \frac{d}{d q^2} (\Pi_V - \Pi_A) \Big|_{q^2 = 0} \, ,
\ee
where $\Pi_V$ and $\Pi_A$ are the vector and axial-vector current two-point functions. It is well-known that the above expression can be rewritten as the following sum over vector and axial-vector resonances:
\be \label{SumRes}
S = 4 \pi \sum_{n} \left( \frac{g_{V_n}^2}{m_{V_n}^4} -\frac{g_{A_n}^2}{m_{A_n}^4} \right) \, .
\ee

Let us assume that we have found a U-shape embedding of probe ${\rm D}p$-$\overline{{\rm D}p}$ branes into the background created by the technicolor ones. To compute the masses and decay constants in (\ref{SumRes}), consider the DBI action of the probe ${\rm D}p$-branes:
\be \label{SDBI}
S_{DBI} = -T \int d^4 x \,d \rho \,d \Omega_{p-4} \,e^{- \phi} \sqrt{- \det (g_{ab} + 2 \pi \alpha' F_{ab})} \,\, ,
\ee
where $a,b = 0,1,...,p$; $g_{ab}$ is the induced metric, $F_{ab}$ is the world-volume (for us, techniflavor) field strength, $\rho$ is the radial direction, $\Omega_{p-4}$ are the compact internal directions wrapped by the ${\rm D}p$ brane and, finally, $\phi$ is the dilaton. To leading (i.e., quadratic) order in $F_{ab}$ this action is:
\be \label{YM}
S_{DBI} = -T \int d^4 x \,d \rho \,d \Omega_{p-4} \,e^{-\phi} \sqrt{- \det (g_{p+1})} \,{\rm Tr} g^{ab} g^{cd} F_{ac} F_{bd} \, . 
\ee
Using the solution for the ${\rm D}p$ profile and integrating over the world volume directions parametrized by $\Omega_{p-4}$, we obtain an action of the form:
\be \label{GFAction}
S_{DBI} = - \frac{\kappa}{4} \int d^4 x \, d \rho \,\left[ a(\rho) F_{\mu \nu} F^{\mu \nu} + 2 b(\rho) F_{\mu \rho} F^{\mu}{}_{\rho} \right] \, ,
\ee
where $\mu, \nu = 0,1,2,3$ and we assume $(-, +, +, ...)$ space-time signature. Furthermore, $\kappa = \frac{T (2 \pi \alpha')^2 V_{p-4}}{g_s}$ with $V_{p-4}$ being the volume of the compact cycle wrapped by the ${\rm D}p$ probe brane and the functions $a(\rho)$ and $b(\rho)$ arise from the $\rho$-dependence of the determinant and of the inverse of the induced metric in (\ref{YM}). 

Now, we want to solve the field equations that follow from the action (\ref{GFAction}). For convenience, we will use the gauge $A_{\rho} (x, \rho) = 0$. Let us Fourier transform the gauge potential $A_{\mu} (x, \rho)$ in the coordinates $x^{\mu}$ and expand:
\be \label{decomp}
A_{\mu} (q, \rho) = {\cal V}_{\mu} (q) \psi_V^0 (q^2, \rho) + {\cal A}_{\mu} (q) \psi_A^0 (q^2, \rho) + \sum_{n} \left( V_{\mu}^n (q) \psi_{V_n} (\rho) + A_{\mu}^n (q) \psi_{A_n} (\rho) \right) \, ,
\ee
where the terms containing $\psi_V^0$ and $\psi_A^0$ are non-normalizable modes, that correspond to sources for the vector and axial-vector boundary currents respectively, whereas the terms in the sum over $n$ are the normalizable modes that correspond to the bulk gauge fields. We have divided these terms into vector and axial-vector ones in the following manner. As in the Sakai-Sugimoto model \cite{SS}, parity in the field theory is related to reflection on the flavor-brane embedding about the point $\rho = \rho_0$ at which the ${\rm D}p$ and $\overline{{\rm D}p}$ stacks join. Hence the vector modes, $\psi_{V_n}$, are those that are symmetric w.r.t. reflection around $\rho_0$ and the axial-vector modes, $\psi_{A_n}$, are those that are anti-symmetric. In order for the four-dimensional action to be canonically normalized, the modes have to satisfy the normalization condition:
\be \label{norm}
\kappa \int d\rho \,\, a(\rho) \, \psi_{V_n} \psi_{V_m} = \delta_{nm}
\ee  
and similarly for $\psi_{A_n}$. The boundary conditions for the normalizable modes, as $\rho \rightarrow \infty$, are $\psi_{V_n}, \psi_{A_n} \rightarrow 0$ on both branches of the U-shaped ${\rm D}p$-$\overline{{\rm D}p}$ world-volume. Also, by definition the symmetric modes satisfy $\pd_{\rho} \psi_{V_n} (\rho)|_{\rho = \rho_0} = 0$, whereas the antisymmetric ones satisfy  $\psi_{A_n} (\rho_0) = 0$. On the other hand, for the non-normalizable modes one has $\psi_V^0 (q^2, \infty) = 1$ on both branches of the U-shape embedding (i.e., both on the stack of ${\rm D}p$'s and on the stack of $\overline{{\rm D}p}$'s), whereas $\psi_A^0 (q^2, \infty) = 1$ on the ${\rm D}p$ branes and $\psi_A^0 (q^2, \infty) = -1$ on the $\overline{{\rm D}p}$ branes.

Substituting the decomposition (\ref{decomp}), one finds that the equations of motion (both for the vector and for the axial-vector modes) that follow from (\ref{GFAction}) are:
\be \label{normFE}
\frac{1}{a(\rho)} \,\pd_{\rho} [b(\rho) \pd_{\rho} \psi_n (\rho)] = - m_n^2 \psi_n (\rho) \, ,
\ee
where we have used that\footnote{The condition $\pd^{\mu} V_{\mu}^n = 0$ follows from varying the action (\ref{GFAction}) with respect to the radial component $A_{\rho}$ and then imposing the $A_{\rho} = 0$ gauge.}
\be
\pd^{\nu} \pd_{\nu} V_{\mu}^n = m_n^2 V_{\mu}^n \qquad {\rm and} \qquad \pd^{\mu} V_{\mu}^n = 0 \, .
\ee 
For the non-normalizable modes, one has the same equations of motion except for the change $m_n^2 \rightarrow q^2$:
\be \label{NNMode}
\frac{1}{a(\rho)} \,\pd_{\rho} [b(\rho) \pd_{\rho} \psi_{V,A}^0 (q^2, \rho)] = - q^2 \psi_{V,A}^0 (q^2, \rho) \, .
\ee

At first sight, it might seem that summing over only the first several resonances in (\ref{SumRes}) would give a reasonable estimate for the S-parameter. This intuition turns out to be incorrect: It was shown in \cite{HolTech} that restricting the infinite sum over resonances to a finite number of the lowest lying ones is, generically, an inaccurate approximation because of non-decoupling of KK modes.\footnote{This is also consistent with the conclusions of \cite{HYS}.} However, in the large-$N_{TC}$ limit one can compute the S-parameter exactly by using the non-normalizable modes. We will go over the derivation of this result in the next subsection and, by being a little more careful than \cite{HolTech}, we will end up with a slightly different formula.

\subsection{Holographic S-parameter formula}

Let us start by recalling that the two-point functions $\Pi_V$ and $\Pi_A$ in (\ref{Sdef}) are related to the vector $J_{\mu}^V$ and axial-vector $J_{\mu}^A$ current correlators in the following manner: 
\bea
i \left( g_{\mu \nu} - \frac{q_{\mu} q_{\nu}}{q^2} \right) \delta^{\hat{a}\hat{b}} \,\Pi_V (q^2) &=& \int d^4 x \,e^{-i q.x} \langle J_{\mu}^{\hat{a} V} (x) J_{\nu}^{\hat{b} V} (0) \rangle \,\,\, , \nn \\
i \left( g_{\mu \nu} - \frac{q_{\mu} q_{\nu}}{q^2} \right) \delta^{\hat{a}\hat{b}} \,\Pi_A (q^2) &=& \int d^4 x \,e^{-i q.x} \langle J_{\mu}^{\hat{a} A} (x) J_{\nu}^{\hat{b} A} (0) \rangle \,\,\, ,
\eea
where $\hat{a}, \hat{b} = 1,...,N_{TF}^2 - 1$ are labeling the techniflavor currents. Now, according to the decomposition (\ref{decomp}), the sources for $J_{\mu}^V$ and $J_{\mu}^A$ are ${\cal V_{\mu}}$ and ${\cal A}_{\mu}$, respectively. Hence, using the gauge/gravity duality (more precisely, the statement that the generating functional of the field theory correlators is given by the dual gravity action), we can compute $\Pi_V$ by:
\be \label{PiHol}
\Pi_V (q^2) = \langle J_{\mu}^V (q^2) J_{\nu}^V (0) \rangle = - \frac{\delta}{\delta {\cal V}_{\mu}} \frac{\delta}{\delta {\cal V}_{\nu}} S_{DBI} \Big|_{{\cal V}=0}
\ee
and similarly for $\Pi_A$.

To utilize (\ref{PiHol}), let us now consider in more detail the action $S_{DBI}$ in (\ref{GFAction}) with (\ref{decomp}) substituted. After Fourier transforming to momentum space $x^{\mu} \rightarrow q^{\mu}$, we find:
\bea \label{SDBIcalc}
S_{DBI} &=& - \frac{\kappa}{4} \int d^4 q \,d\rho \,{\rm Tr} \bigg\{ a(\rho) \bigg( \sum_n \left[ \,|F_{\mu \nu}^{V_n} (q)|^2 \,\psi^2_{V_n} \!(\rho) + |F_{\mu \nu}^{A_n} (q)|^2 \,\psi^2_{A_n} \!(\rho) \right. \bigg. \bigg. \nn \\
&+& \left. 2 F_{\mu \nu}^{{\cal V}} (q) F^{\mu \nu}_{V_n} (-q) \,\psi_V^0 \!(\rho) \,\psi_{V_n} \!(\rho) + 2 F_{\mu \nu}^{{\cal A}} (q) F^{\mu \nu}_{A_n} (-q) \,\psi_A^0 \!(\rho) \,\psi_{A_n} \!(\rho) \,\right] \nn \\
&+& \bigg. |F_{\mu \nu}^{{\cal V}} (q)|^2 (\psi^0_V (\rho))^2 + |F_{\mu \nu}^{{\cal A}} (q)|^2 (\psi^0_A (\rho))^2 \bigg) \!+ 2 b(\rho) \bigg( |{\cal V}_{\mu} (q)|^2 (\pd_{\rho} \psi_V^0)^2 \bigg. \nn \\
&+& |{\cal A}_{\mu} (q)|^2 (\pd_{\rho} \psi_A^0)^2 + \sum_n \left[ \,|V_{\mu}^n (q)|^2 (\pd_{\rho} \psi_{V_n})^2 + |A_{\mu}^n (q)|^2 (\pd_{\rho} \psi_{A_n})^2 \right. \nn \\
&+& \bigg. \bigg. \left. 2 {\cal V}_{\mu} V^{\mu}_n (\pd_{\rho} \psi^0_V) (\pd_{\rho} \psi_{V_n}) + 2 {\cal A}_{\mu} A^{\mu}_n (\pd_{\rho} \psi^0_A) (\pd_{\rho} \psi_{A_n}) \,\right] \bigg) \bigg\} \, .
\eea
Now let us use (\ref{norm}) in the first line of (\ref{SDBIcalc}). In addition, let us substitute $\psi_{V_n,A_n}$ in the second line with the corresponding expression from (\ref{normFE}), i.e. $\psi_{n} = - \frac{1}{m^2_{n}} \frac{1}{a} \pd_{\rho} (b \pd_{\rho} \psi_{n})$. Also, let us partially integrate the $\pd_{\rho} \psi_{V_n,A_n}$ terms inside the bracket multiplying $b(\rho)$ and then use (\ref{normFE}), (\ref{NNMode}). As a result of these manipulations, we obtain:
\bea \label{SDBI2}
S_{DBI} &=& - {\rm Tr} \int d^4 q \sum_n \bigg( \frac{1}{4} |F_{\mu \nu}^{V_n} (q)|^2 + \frac{1}{4} |F_{\mu \nu}^{A_n} (q)|^2 + \frac{1}{2} m_{V_n}^2 |V_{\mu}^n (q)|^2 + \frac{1}{2} m_{A_n}^2 |A_{\mu}^n (q)|^2 \bigg. \nn \\
&+& \bigg. a_{V_n} F_{\mu \nu}^{{\cal V}} (q) F_{V_n}^{\mu \nu} (q) + a_{A_n} F_{\mu \nu}^{{\cal A}} (q) F_{A_n}^{\mu \nu} (q) \bigg) + S_{source} \,\, ,
\eea
where
\be
a_{V_n} = - \frac{\kappa}{m_{V_n}^2} \int d \rho \,\, \psi_V^0 (\rho) \,\pd_{\rho} \!\left[ b(\rho) \psi_{V_n} \!(\rho) \right] \quad , \quad a_{A_n} = - \frac{\kappa}{m_{A_n}^2} \int d \rho \,\, \psi_A^0 \,\pd_{\rho} \!\left[ b(\rho) \psi_{A_n} \!(\rho) \right]
\ee
and $S_{source}$ is the term that contains only the sources. More precisely, we have:
\bea \label{Ssource}
S_{source} &=& - \frac{\kappa}{4} \int d^4 q \,d\rho \,{\rm Tr} \bigg( 2 b(\rho) \bigg[ |{\cal V}_{\mu} (q)|^2 (\pd_{\rho} \psi_V^0)^2 + |{\cal A}_{\mu} (q)|^2 (\pd_{\rho} \psi_A^0)^2 \bigg] \bigg. \nn \\
&+& \bigg. a(\rho) \bigg[ |F_{\mu \nu}^{\cal V} (q)|^2 (\psi^0_V (\rho))^2 + |F_{\mu \nu}^{\cal A} (q)|^2 (\psi^0_A (\rho))^2 \bigg] \bigg) \nn \\
&=& - \frac{1}{2} \,{\rm Tr} \int d^4 q \,\bigg( a_V^0 (q) |{\cal V}_{\mu} (q)|^2 + a_A^0 (q) |{\cal A}_{\mu} (q)|^2 \bigg) \, ,
\eea
where 
\bea
a_V^0 (q^2) &=& 2 \kappa \left[ b(\rho) \,\psi_V^0 (q^2, \rho) \,\pd_{\rho} \psi_V^0 (q^2, \rho) \right]_{\rho = \infty} \, , \nn \\
a_A^0 (q^2) &=& 2 \kappa \left[ b(\rho) \,\psi_A^0 (q^2, \rho) \,\pd_{\rho} \psi_A^0 (q^2, \rho) \right]_{\rho = \infty}
\eea
with the factor of 2 being due to the two branches of the D7-$\overline{{\rm D}7}$ and, also, to obtain the last line in (\ref{Ssource}) we have integrated by parts the $(\pd_{\rho} \psi_{V,A}^0)^2$ terms on the first line there. To cancel the second line in $S_{source}$ we have also used that $|F_{\mu \nu}^{\cal V} (q)|^2 = - 2 q^2 |{\cal V}_{\mu} (q)|^2$, which is due to the Fourier mapping $\pd_{\mu} \rightarrow i q_{\mu}$. As in \cite{HolTech,SS2}, one can diagonalize the kinetic terms in (\ref{SDBI2}) by introducing:
\be
\tilde{V}_{\mu}^n = V_{\mu}^n + a_{V_n} {\cal V}_{\mu} \,\, , \qquad \tilde{A}_{\mu}^n = A_{\mu}^n + a_{A_n} {\cal A}_{\mu} \,\, .
\ee
Then, from the coupling of the new fields $\tilde{V}_{\mu}^n$ and $\tilde{A}_{\mu}^n$ with the sources, one can read off the decay constants:
\be \label{decayConst}
g_{V_n} = m_{V_n}^2 a_{V_n} = \kappa \int d \rho \, \psi_V^0 \,\psi_{V_n} \,\, , \qquad g_{A_n} = m_{A_n}^2 a_{A_n} = \kappa \int d\rho \, \psi_A^0 \,\psi_{A_n} \,\, .
\ee

Now, using (\ref{PiHol}) we obtain:
\bea \label{PiVA}
\Pi_V (q^2) \!\!&=& \!\! a_V^0 (q^2) \,= 2\,\kappa \left[ b(\rho) \,\psi_V^0 (q^2, \rho) \,\pd_{\rho} \psi_V^0 (q^2, \rho) \right]_{\rho = \infty} \,\, , \nn \\
\Pi_A (q^2) \!\!&=& \!\! a_A^0 (q^2) \,= 2\,\kappa \left[ b(\rho) \,\psi_A^0 (q^2, \rho) \,\pd_{\rho} \psi_A^0 (q^2, \rho) \right]_{\rho = \infty} \,\, .
\eea
Therefore, (\ref{Sdef}) implies that
\be \label{Sexact}
S = - 8 \pi \kappa \left[ b(\rho) \,\frac{\pd}{\pd q^2} \!\left( \psi_V^0 \pd_{\rho} \psi_V^0 - \psi_A^0 \pd_{\rho} \psi_A^0 \right) \right]_{\rho = \infty, \,q^2 = 0} \, .
\ee
Note that this expression is slightly different from the one derived and used in \cite{HolTech}, which is $S = -4 \pi \kappa \left[ b(\rho) \frac{\pd}{\pd q^2} \left( \pd_{\rho} \psi_V^0 - \pd_{\rho} \psi_A^0 \right) \right]_{\rho = \infty, \,q^2 = 0}$. The reason for the discrepancy (other than the overall factor of two) is that the authors of \cite{HolTech} have substituted the boundary condition $\psi^0_{V,A} (\rho = \infty) = 1$ in the intermediate steps of the computation. However, on general grounds it should be clear that, by performing this substitution before taking the limit $\rho \rightarrow \infty$, one can miss some of the contributions to $S$. Indeed, we will see below that this is precisely what happens in the holographic technicolor models.

Before concluding this section, let us make one more remark. Clearly, a rescaling $\psi^0_{V,A} \rightarrow C \psi^0_{V,A}$, with $C$ an arbitrary constant, does not spoil the solution of (\ref{NNMode}). As in \cite{SS2}, under such a rescaling the decay constants rescale as $g_n \rightarrow C g_n$, according to (\ref{decayConst}). This then implies, due to (\ref{SumRes}), that the S-parameter rescales as $S \rightarrow C^2 S$, which is also consistent with (\ref{Sexact}). In our context, this freedom of rescaling is fixed by imposing that $\Pi_A (0) = F_{\pi}^2$, where $F_{\pi} = 250 \,{\rm GeV}$ is the technipion decay constant.\footnote{Recall that $250 \,{\rm GeV}$ is roughly the electroweak scale.} 

\section{Regular technicolor: example} \label{RegTechEx}
\setcounter{equation}{0}

In \cite{HolTech}, the expression for the S-parameter, whose derivation we reviewed above, was evaluated numerically by solving numerically the field equations for $\psi_{V,A}^0$. Instead of doing that for our improved formula, we will try to gain more insight by analyzing things analytically. Before turning to the walking background of \cite{NPP}, let us first reconsider one of the regular technicolor models of the last reference in \cite{HolTech} as an example of our approach. Namely, we will look at the model obtained by placing D7-$\overline{\rm D7}$ probes in the conifold.

\subsection{D-brane set-up}

As shown in \cite{KS}, one can have a geometric realization of chiral symmetry breaking by embedding D7-$\overline{\rm D7}$ flavor branes in the conifold, as that embedding has the characteristic U-shape profile. To be able to be more precise, let us first briefly recall a few things about the conifold geometry. The 10d metric is given by
\be \label{Con10dM}
ds^2_{10} = \frac{r^2}{R^2} dx^2 + \frac{R^2}{r^2} ds^2_{6} \,\, ,
\ee  
where $x^{\mu}$ are 4d coordinates and 
\be
ds^2_{6} = dr^2 + \frac{r^2}{3} \left( \frac{1}{4} (f_1^2 + f_2^2) + f_3^2 + (d \theta - \frac{1}{2} f_2)^2 + (\sin \theta d \varphi -\frac{1}{2} f_1)^2 \right)
\ee
with the one-forms $f_i$ parameterizing a three-sphere. Now, let us choose a D7 embedding such that the transverse space is the two-sphere parameterized by the angular coordinates $\theta$ and $\varphi$. In other words, the D7-brane worldvolume is spanned by $\{ x^{\mu} \}$, $r$ and $\{ f_i \}$. To specify completely the embedding of the eight-dimensional worldvolume into the ten-dimensional space-time, we also need an ansatz for the position of the brane in the transverse space. So let us assume that $\theta$ and $\varphi$ depend only on the radial variable $r$. Then, substituting $\theta = \theta (r)$ and $\varphi = \varphi (r)$ into (\ref{Con10dM}) in order to obtain the metric $g_{8}$ induced on the worldvolume, the D7-brane DBI action
\be
S_{DBI} = - \mu_7 \int e^{-\phi} \sqrt{-\det(g_{8})} 
\ee
leads to the Lagrangian \cite{KS}:
\be \label{ConD7Lag}
{\cal L} =  - \mu_7 \,e^{-\phi} \,\frac{r^3}{18} \left( 1 + \frac{r^2}{6} \left( \theta_r^2 + \sin^2 \theta \varphi_r^2 \right) \right)^{1/2} ,
\ee
where we have denoted $\theta_r \equiv \pd \theta / \pd r$ and $\varphi_r \equiv \pd \varphi / \pd r$. Note also that in this background the dilaton $\phi$ is constant. So, as shown in \cite{KS}, the field equations that follow from (\ref{ConD7Lag}) are solved for $\theta = \pi /2$ and $\varphi(r)$ satisfying
\be \label{ConD7emb}
\cos \left( \frac{4}{\sqrt{6}} \varphi (r) \right) = \left( \frac{r_0}{r} \right)^4 \, ,
\ee
where $r_0$ is an integration constant and the other integration constant has been set to zero. One can easily see that for generic $r> r_0$ the solution represents two points on the equator of the  $S^2$ parameterized by $(\theta , \varphi)$; this corresponds to the two separate stacks of D7 and $\overline{\rm D7}$ branes. At $r=r_0$ these two points coincide, which corresponds to the merging of the D7s and anti-D7s.

This model of chiral symmetry breaking was used in Section 9 of the last reference in \cite{HolTech}, in order to obtain a technicolor model according to the general discussion in our Section \ref{HolTechSec}. Using the embedding (\ref{ConD7emb}), one can compute that the coefficient functions $a$ and $b$ in (\ref{GFAction}) acquire the form:
\be \label{abz}
a (z) = \frac{1}{\sqrt{z^2 + \left(\frac{r_0}{R}\right)^8 }} \qquad {\rm and} \qquad b(z) = 16 \left( z^2 + \left(\frac{r_0}{R}\right)^8 \right)^{3/2} ,
\ee
where $z \in (-\infty, +\infty)$ is a suitably chosen worldvolume variable, such that $z>0$ runs along the D7 branch and $z<0$ runs along the $\overline{\rm D7}$ branch, unlike the space-time radial variable $r\in (r_0, \infty)$ that does not distinguish between the two branches. More precisely, $z^2 = \frac{r^2}{R^2} \left( 1- \frac{r_0^8}{r^8} \right)$; see \cite{KS}.
 
Now, instead of solving numerically the field equations
\be \label{FEz}
\frac{1}{a(z)} \, \pd_z \!\left[ b(z) \pd_z \psi^0_{V,A} (q^2, z) \right] \!= - q^2 \psi^0_{V,A} (q^2, z) \,\, ,
\ee 
as done in \cite{HolTech}, we will try to determine analytically the behaviour of the functions $\psi^0_{V,A}$ in order to evaluate the S-parameter. Although we will be left with several undetermined integration constants, whose values can only be fixed by a numerical computation, our considerations will illuminate some interesting/important points.
 
\subsection{S-parameter}

Although equation (\ref{FEz}) cannot be solved analytically in the whole domain of variation of $z$, it can be solved analytically for $z >\!\!> \frac{r_0}{R}$, which is precisely the region of interest in the evaluation of the S-parameter formula (\ref{Sexact}). Namely, for large $z$ (\ref{FEz}) acquires the form:\footnote{Here we also take $z>0$. Since (\ref{Sexact}) already takes into account that the contribution of the two branches is equal to twice the contribution of only one of them, in the following we will concentrate only on the D7 branch.}
\be \label{FEzs}
16 z^4 \frac{\pd^2}{\pd z^2} \psi^0_{V,A} + 48 z^3 \frac{\pd}{\pd z} \psi^0_{V,A} + q^2 \psi^0_{V,A} = 0 \, .
\ee
The latter equation is solved by
\be \label{ExSol}
\psi^0_{V,A} = C_1^{V,A} \,\frac{1}{z} \,J_1 \!\left( \frac{q}{4z} \right) + C_2^{V,A} \,\frac{1}{z} \,Y_1 \!\left( \frac{q}{4z} \right) ,
\ee
where $J$ and $Y$ are the Bessel functions of the first and second kind respectively. Note also that the integration constants $C_{1,2}^{V,A}$ can, in principle, depend on $q$ as the latter is just a parameter in the differential equation (\ref{FEzs}).

Now, it may seem that one can just plug the solution (\ref{ExSol}) in the formula (\ref{Sexact}) and obtain a finite answer. However, a direct substitution leads to either zero or infinity, depending on whether one takes the $J_1$ or the $Y_1$ term in the solution. Furthermore, since in the region of interest for us, namely for $z\rightarrow \infty$ and $q\rightarrow 0$, we have that $\frac{1}{z} \,J_1\! \left( \frac{q}{4z} \right) \rightarrow 0$ and $\frac{1}{z} \,Y_1 \!\left( \frac{q}{4z} \right) \rightarrow \infty$, none of the terms in (\ref{ExSol}) tends to a constant and so the boundary condition $\psi^0 (z=\infty) = 1$, used in \cite{HolTech}, cannot be imposed. To overcome all of these problems, we need to somehow 'regulate' the terms in the solution (\ref{ExSol}). In other words, we need to find the analogue of more suitable ('regular') "coordinates" for the present case. 

The idea for how to achieve the desired 'regulation' comes from the original AdS/CFT correspondence. Recall that, as noticed in \cite{FMMR}, generic 2-point correlation functions for scalars in AdS do not approach a constant as one goes toward the boundary. Instead, they tend to zero or infinity. In particular, the solution of the wave equation $ (\nabla^2 - m^2) \phi = 0$ for a scalar with mass $m$ is given by the modified Bessel function of the second kind. More precisely, in momentum space one has:
\be
\phi \sim z^{\frac{d}{2}} K_{\nu} (qz) \phi_0 (q) \, ,
\ee
where $z$ is the radial variable for the ${\rm AdS}_{d+1}$ metric in Poincar\'{e} coordinates, i.e. $ds^2 = \frac{R^2}{z^2} (dz^2 - dt^2 + dx_{d-1}^2)$ so that the boundary is reached for $z \rightarrow 0$ and the deep interior for $z$ large; $\phi_0 (q)$ is a function of the boundary momentum $q$ and $\nu = \sqrt{\frac{d^2}{4}+m^2}$. Now, for $z \rightarrow 0$ the expression $z^{\frac{d}{2}} K_{\nu} (qz)$ diverges.\footnote{This statement applies for $m^2>0$. Recall that in AdS stability does not require positive $m^2$, just that the latter satisfy the Breitenlohner-Freedman bound.} To regulate it, the authors of \cite{FMMR} introduced a cut-off $\epsilon > 0$ and rescaled the solution in the following manner:
\be
\phi = \frac{z^{\frac{d}{2}} K_{\nu} (qz)}{\epsilon^{\frac{d}{2}} K_{\nu} (q \epsilon)} \phi_0 (q) \, ,
\ee
so that $\phi \rightarrow \phi_0 (q)$ for $z \rightarrow \epsilon$. As shown in \cite{FMMR}, using this bulk solution in the supergravity action and performing the computation of the CFT correlators at $z = \epsilon$, before taking the $\epsilon \rightarrow 0$ limit, is the appropriate procedure to extract the correct field theory correlation functions from the gravity dual. 

We would like to adopt the above procedure for our case, in order to obtain finite limits from the $J_1$ and $Y_1$ terms as $q \rightarrow 0$. However, there is an important subtlety. Namely, in the above paragraph the divergence was occurring in a limit of the variable of the differential equation one is solving. In our case, on the other hand, $q$ is just a parameter in the differential equation of interest and the variable is $z$. So we cannot simply rescale, say, $Y_1 \!\left( \frac{q}{4z} \right) \rightarrow Y_1 \!\left( \frac{q}{4z} \right) \!/ Y_1 \!\left( \frac{\epsilon}{4z} \right)$ without spoiling the solution of our differential equation. However, we are allowed to do the following rescaling:
\be \label{expCon}
\frac{1}{z} \,Y_1 \!\left( \frac{q}{4z} \right) \rightarrow \frac{\frac{1}{z} \,Y_1 \!\left( \frac{q}{4z} \right)}{\frac{1}{z_*} \,Y_1 \!\left( \frac{\epsilon}{4z_*} \right)} \, ,
\ee
where $z_*$ is some finite fixed value. In fact, it is not even necessary to introduce a lower bound on the range of variation of $q$, since the right-hand side of (\ref{expCon}) with $\epsilon = q$ is well-behaved in the limit $q \rightarrow 0$. Namely, using the small argument expansion of the $Y_1$ Bessel function, we find:
\be \label{ConY}
\frac{\frac{1}{z} \,Y_1 \!\left( \frac{q}{4z} \right)}{\frac{1}{z_*} \,Y_1 \!\left( \frac{q}{4z_*} \right)} = 1 + C_Y^1 q^2 - C_Y^2 \frac{q^2}{z^2} + {\cal O} (q^4) \, ,
\ee
where
\be
C_Y^1 = \frac{2 \gamma - 1}{64 z^2_*} + \frac{1}{32 z_*^2} \ln \left( \frac{q}{8 z_*} \right) \, , \qquad C_Y^2 = \frac{2 \gamma - 1}{64} + \frac{1}{32} \ln \left( \frac{q}{8 z} \right) \, .
\ee
Similarly, we can compute:
\be \label{ConJ}
\frac{\frac{1}{z} \,J_1 \!\left( \frac{q}{4z} \right)}{\frac{1}{z_*} \,J_1 \!\left( \frac{q}{4z_*} \right)} = \frac{z_*^2}{z^2} + \frac{q^2}{128 z^2} \left( 1 - \frac{z_*^2}{z^2} \right) + {\cal O} (q^4) \, .
\ee
Hence, the solution we are looking for is a linear combination of (\ref{ConY}) and (\ref{ConJ}) with coefficients that may still depend on $q$. We will fix this dependence by finding the small $q$ solution of (\ref{FEzs}) in yet another, more direct, manner.

Namely, a more direct way of solving (\ref{FEzs}) for small $q$ is the following. Let us first consider the equation:
\be
16 z^4 \frac{\pd^2}{\pd z^2} \psi^0 + 48 z^3 \frac{\pd}{\pd z} \psi^0 = 0 \,\, .
\ee
Its most general solution is $f_1(q) + f_2 (q) / z^2$, where $f_1 (q)$ and $f_2 (q)$ are arbitrary functions of $q$. Now, let us recall that we are looking for a solution that is an expansion in small $q^2$ and tends to 1. Then the above general solution reduces to:
\be
\psi^0_h = 1+ \tilde{C}_1 q^2 + \frac{(\tilde{C}_2 + \tilde{C}_3 q^2)}{z^2} + {\cal O} (q^4) \,\, ,
\ee 
where $\tilde{C}_1$, $\tilde{C}_2$ and $\tilde{C}_3$ are constants and we have stopped at ${\cal O} (q^2)$ since terms of ${\cal O} (q^4)$ and higher do not contribute to the expression for the S-parameter (\ref{Sexact}). Now, we can solve (\ref{FEzs}) to order $q^2$ by adding to $\psi_h^0$ a particular solution of the inhomogeneous equation, obtained by substituting in the last term of (\ref{FEzs}) the zeroth order of $\psi_h^0$. Namely, the inhomogeneous equation of interest is:
\be
16 z^4 \frac{\pd^2}{\pd z^2} \psi^0 + 48 z^3 \frac{\pd}{\pd z} \psi^0 + \,q^2 \!\left( 1 + \frac{\tilde{C}_2}{z^2} \right) \!= \,0 \,\, .
\ee
It is solved by:
\be
\psi^0_i = \frac{q^2 \ln z}{32 z^2} + \frac{q^2}{64 z^2} - \frac{q^2 \tilde{C}_2}{128 z^4} \, ,
\ee
where the integration constants have been set to zero since their non-vanishing contributions are already taken into account within the constants in $\psi_h^0$. Hence the solution of (\ref{FEzs}) to order $q^2$ is:
\be \label{ConSmallqSol}
\psi^0_{V,A} = \psi_h^0 + \psi_i^0 = 1+ \tilde{C}_1^{V,A} q^2 + \frac{(\tilde{C}_2^{V,A} + \tilde{C}_3^{V,A} q^2)}{z^2} + \frac{q^2}{32 z^2} \!\left( \ln z + \frac{1}{2} - \frac{\tilde{C}_2^{V,A}}{4 z^2} \right) \!+ {\cal O} (q^4) \, .
\ee 
Using the above solution, together with (\ref{abz}) for large $z$, we can compute that:
\be \label{ingred}
b(z) \,\pd_{q^2} \!\!\left( \psi^0 \pd_z \psi^0 \right)\Big|_{q^2 = 0} \!= - \ln z - 32 \left( \tilde{C}_3 + \tilde{C}_1 \tilde{C}_2 \right) - \frac{\tilde{C}_2}{z^2} \left( 64 \tilde{C}_3 + 2 \ln z - \frac{3 \tilde{C}_2}{4 z^2} \right) \, .
\ee
Therefore, from (\ref{Sexact}) we find:
\be
S = 256\,\pi \kappa \left( \tilde{C}_3^V + \tilde{C}_1^V \tilde{C}_2^V - \tilde{C}_3^A - \tilde{C}_1^A \tilde{C}_2^A \right) \, .
\ee
 
Despite still having to determine the values of the constants $\tilde{C}_{1,2,3}$ numerically, our considerations so far enable us to make several important observations. First of all, note that the leading term in (\ref{ingred}), namely $\ln z$, is divergent for $z \rightarrow \infty$. However, it is the same for $V$ and $A$ modes and thus cancels in the S-parameter expression. Hence, the S-parameter results from very small differences between large $V$ and $A$ contributions. Another important point is that, if instead of (\ref{Sexact}) we had used $S = -4 \pi \kappa \left[ b(\rho) \frac{\pd}{\pd q^2} \left( \pd_{\rho} \psi_V^0 - \pd_{\rho} \psi_A^0 \right) \right]_{\rho = \infty, \,q^2 = 0}$ as in \cite{HolTech}, we would have missed the $\tilde{C}_1 \tilde{C}_2$ contribution.\footnote{In fact, only the $\tilde{C}^A_1 \tilde{C}^A_2$ term in $\tilde{C}^V_1 \tilde{C}^V_2 - \tilde{C}^A_1 \tilde{C}^A_2$ will contribute, since $\Pi_V(0) = 0$ implies that $\tilde{C}_2^V = 0$.} It is also worth noting that the initial intuition one might have had, namely to drop the $Y_1$ term in the solution (\ref{ExSol}) as it diverges for $q \rightarrow 0$, is actually incorrect. Indeed, comparing (\ref{ConSmallqSol}) with (\ref{ConY}) and (\ref{ConJ}), one can see that the $Y_1$ term has an essential contribution to the final answer.

To compare in more detail the small $q^2$ solution (\ref{ConSmallqSol}) with the regulated version of the general $q$ solution $C_1 J_1/z + C_2 Y_2/z$, let us note that there is no reason to have the same value $z_*$ in the regulation of both the $J_1$ and the $Y_1$ terms. In other words, we can have $z_*^J \neq z_*^Y$. This is important, since the small $q$ solution (\ref{ConSmallqSol}) has three independent constants $\tilde{C}_{1,2,3}$. Therefore, the small $q$ expansion of $C_1 J_1/z + C_2 Y_2/z$ should have three constants as well. Since the constant piece in (\ref{ConSmallqSol}) has been normalized to 1, then from (\ref{ConY}) it follows that $C_2 = 1$. Hence the three independent constants are actually $C_1$, $z_*^J$ and $z_*^Y$. Note also that, unlike (\ref{ConSmallqSol}), the expansion in (\ref{ConY}) has terms of the form $q^2 \ln q$. We can cancel the $q^2 \ln q$ term in $C_Y^1$ by multiplying (\ref{ConY}) by $(1 - \frac{1}{32 (z_*^Y)^2} q^2 \ln q)$ and the $q^2 \ln q$ term in $C_Y^2$ by multiplying (\ref{ConJ}) by $(1 + \frac{1}{C_1 32 (z_*^J)^2} q^2 \ln q)$; these manipulations do not affect the rest of the ${\cal O} (q^2)$ expansions. To recapitulate, the small $q$ solution (\ref{ConSmallqSol}) arises from the small $q$ expansion of the solution (\ref{ExSol}) for the following choice of the integration constants:
\be
\psi^0 (q^2, z) = \frac{C_1 \!\left(1+ \frac{q^2 \ln q}{32 C_1 (z_*^J)^2} \right)}{\frac{1}{z_*^J} \,J_1 \!\left( \frac{q}{4 z_*^J} \right)} \,\,\frac{1}{z} \,J_1 \!\left( \frac{q}{4z} \right) \,+ \,\frac{\left(1- \frac{q^2 \ln q}{32 (z_*^Y)^2} \right)}{\frac{1}{z_*^Y} \,Y_1 \!\left( \frac{q}{4 z_*^Y} \right)} \,\,\frac{1}{z} \,Y_1 \!\left( \frac{q}{4z} \right) \, .
\ee
 
\section{Walking technicolor} \label{WalkT}
\setcounter{equation}{0}

In this section we study our model of walking technicolor. First, we review briefly the background of \cite{NPP}, that is dual to walking behaviour. Then we show that one can realize geometrically chiral symmetry breaking in it, i.e. that there is a U-shape embedding a la Sakai-Sugimoto of ${\rm D}7$-$\overline{{\rm D}7}$ probes in this background. Using that result, we then compute the S-parameter with the method, illustrated in the previous section. The answer turns out to have divergences, which we remove by adding an appropriate counterterm. It is, perhaps, interesting to note that the counterterm also gives a finite contribution to the renormalized S-parameter.

\subsection{Gravity background} \label{GrBackground}

The gravitational background that we will consider is given in eq. (6) of \cite{NPP}. Namely, the ten-dimensional string frame metric is: 
\bea \label{bckgrM}
\hspace*{-0.8cm}ds^2 &=& \alpha' g_s e^{\phi(\rho)} \left[ \frac{dx^2_{1,3}}{\alpha' g_s} + e^{2 k(\rho)} d\rho^2 + e^{2 h(\rho)} (d\theta^2 + \sin^2 \theta d\varphi^2) \right. \nn \\
\hspace*{-0.8cm}&+& \left. \frac{e^{2 g(\rho)}}{4} \{ (\tilde{\omega}_1 + a(\rho) d\theta )^2 + ( \tilde{\omega}_2 - a(\rho) \sin \theta d\varphi )^2 \} 
+ \frac{e^{2 k(\rho)}}{4} (\tilde{\omega}_3 + \cos \theta d\varphi)^2 
\right],
\eea
where 
\bea
\tilde{\omega}_1 &=& \cos \psi d\tilde{\theta} + \sin \psi \sin \tilde{\theta} d \tilde{\varphi} \nn \\
\tilde{\omega}_2 &=& - \sin \psi d\tilde{\theta} + \cos \psi \sin \tilde{\theta} d \tilde{\varphi} \nn \\
\tilde{\omega}_3 &=& d \psi + \cos \tilde{\theta} d \tilde{\varphi}
\eea
and the functions $\phi(\rho)$, $k(\rho)$, $h(\rho)$, $g(\rho)$, and $a(\rho)$ are  determined by the type IIB equations of motion.\footnote{There are no nice analytic expressions for those functions in general. However, they can all be expressed via the BPS conditions in a compact way in terms of a single function, that satisfies certain second order differential equation; see \cite{HBNP} for details.} Note that there is also a nonzero $F_3$ flux. However, its explicit form will not be of importance for us in the following. As in \cite{NPP}, we will take from now on $\alpha' g_s =1$. Finally, an important property of this IIB solution is that the dilaton $\phi(\rho) = const$ \cite{NPP}.

In \cite{NPP} it was shown that there is an intermediate region for the radial variable $\rho$, such that the metric simplifies to: 
\be \label{simpM}
ds^2 \approx \frac{\sqrt{3}}{c^{3/2} \sin^{3/2} \alpha} \left[ dx^2_{1,3} + \frac{c \cos \alpha}{4} \left( \frac{\tan^3 \alpha \,e^{4 \rho}}{3} \left( 4 d \rho^2 + (\omega_3 + \tilde{\omega}_3)^2 \right) + d\Omega^2_2 + d\tilde{\Omega}^2_2 \right) \right],
\ee
where $c$ and $\alpha$ are constants, whereas $d\Omega^2_2 = \omega^2_1 + \omega^2_2$ and  $d\tilde{\Omega}^2_2 = \tilde{\omega}_1^2 + \tilde{\omega}_2^2$, and finally
\be 
\omega_1 = d \theta \, , \qquad \omega_2 = \sin \theta d\varphi \, , \qquad \omega_3 = \cos \theta d\varphi \, . 
\ee
This region corresponds in the dual field theory to the intermidiate energy range, in which the gauge coupling is approximately constant. In other words, it corresponds exactly to the walking regime we are interested in.

Another useful limit, in which (\ref{bckgrM}) simplifies significantly, is the UV region. In this case, the metric is \cite{NPP}:
\be \label{UVM}
ds^2 \approx \frac{\sqrt{3}}{c^{3/2} \sin^{3/2} \alpha} \left[ dx^2_{1,3} + \frac{2^{-1/3} c \sin \alpha}{4} \,\, e^{4 \rho / 3} \left( \frac{2}{3} \left( 4 d \rho^2 + (\omega_3 + \tilde{\omega}_3)^2 \right) + d\Omega^2_2 + d\tilde{\Omega}^2_2 \right) \right].
\ee 

\subsection{D7-brane probes}

We will introduce techniflavors in the above background via D7 probes. Our goal will be to find a geometric realization of chiral symmetry breaking a la Sakai-Sugimoto \cite{SS}. Then one could realize the holographic description of walking technicolor in the manner we reviewed in Section 2.

We take the D7 branes to span the dimensions parametrized by the following coordinates: the four space-time directions $x^{\mu}$, the radial direction $\rho$ and the triplet ($\psi$, $\tilde{\theta}$, $\tilde{\varphi}$), which parameterizes a 3-sphere. Hence, the transverse space is the two-sphere parameterized by ($\theta$, $\varphi$). As reviewed in Section 3, this embedding is of the same kind as the embedding of D7 probes in the conifold, considered in \cite{KS}. Following that paper, we assume that $\theta$ and $\varphi$ depend only on $\rho$. Therefore, the induced metric on the D7 world-volume is given by (\ref{bckgrM}) with $d\theta = \frac{\pd \theta}{\pd \rho} d\rho$ and $d\varphi = \frac{\pd \varphi}{\pd \rho} d\rho$ substituted. 

Now, the DBI action is:
\be
S_{D7} = - \mu_7 \int e^{-\phi} \sqrt{-\det(g_8)} \, ,
\ee
where $g_8$ is the induced metric on the 8-dimensional world-volume. Note that the Chern-Simons term does not contribute since the only background flux is $F_3$. Taking $\theta = \theta (\rho)$ and $\varphi = \varphi (\rho)$ into account in (\ref{bckgrM}), we can compute that
\be
\det(g_8) = - \frac{e^{8 \phi + 4 g + 2 k}}{64} \left[ e^{2 h} \left( \theta^2_{\rho} + \sin^2 \theta \varphi^2_{\rho} \right) + e^{2 k} \right],
\ee
where $\theta_{\rho} \equiv \pd \theta / \pd \rho$ and $\varphi_{\rho} \equiv \pd \varphi / \pd \rho$. Hence we have the following Lagrangian:
\be \label{Lag}
{\cal L} = \frac{\mu_7}{8} \,\, e^{3 \phi(\rho) + 2 g(\rho) + k(\rho)} \left( e^{2 k(\rho)} + e^{2 h(\rho)} (\theta^2_{\rho} + \sin^2 \theta \varphi^2_{\rho}) \right)^{1/2}.
\ee
In principle, one can find the allowed D7 embeddings by finding the solutions of the equations of motion for $\theta (\rho)$ and $\varphi (\rho)$, that follow from (\ref{Lag}). In practice, however, the explicit functions $\phi(\rho)$, $g(\rho)$, $k(\rho)$ and $h(\rho)$ are rather involved and at this point it is not clear whether one can find a solution in full generality.

So let us now consider in turn the two simplified metrics (\ref{simpM}) and (\ref{UVM}). We start with the intermediate region, i.e. with (\ref{simpM}). In this case we find:
\be \label{DetSM}
\det (g_8) = - A^8 C^4 B \, e^{4 \rho} \left( 4 B e^{4 \rho} + \theta^2_{\rho} + \sin^2 \theta \varphi^2_{\rho} \right) \, ,
\ee 
where for convenience we have denoted $C = \frac{c \cos \alpha}{4}$, $B = \frac{\tan^3 \alpha}{3}$ and $A = \frac{\sqrt{3}}{c^{3/2} \sin^{3/2} \alpha}$. Since the dilaton $\phi (\rho)$ is constant \cite{NPP}, (\ref{DetSM}) implies that the resulting Lagrangian is:
\be
{\cal L} = const \times e^{2 \rho} \left( 4 B e^{4 \rho} + \theta^2_{\rho} + \sin^2 \theta \varphi^2_{\rho} \right)^{1/2} \, .
\ee
As in \cite{KS}, one can easily verify that the $\theta$ equation of motion is identically satisfied for $\theta = \pi /2$. In fact, any of $\theta = 0\,, \pm \pi / 2\,, \pi$ is a solution, since $\frac{\pd {\cal L}}{\pd \theta_{\rho} } \sim \theta_{\rho}$ and $\frac{\pd {\cal L}}{\pd \theta} \sim \sin \theta \cos \theta$. For convenience, we will take $\theta = \pi / 2$. Then the $\varphi$ equation of motion becomes:
\be
\varphi^3_{\rho} + 2 B e^{4 \rho} \varphi_{\rho \rho} = 0 \, .
\ee
This is solved by 
\be \label{phiEq}
\tanh \left( \frac{\varphi(\rho)}{\sqrt{B} e^{2 \rho_0}} \right) = \pm \sqrt{1-\frac{e^{4 \rho_0}}{e^{4 \rho}}} \,\, ,
\ee
where we have taken one of the two integration constants to be zero and the other one (up to a constant containing $B$) is denoted by $\rho_0$. Clearly, for a generic value of $\rho$, satisfying $\rho > \rho_0$, there are two solutions of (\ref{phiEq}), which represent two points on the equator of the two-sphere parametrized by $(\theta, \varphi)$. At $\rho = \rho_0$ these two points coincide as the equation $\tanh \varphi = 0$ has the single solution $\varphi = 0$. So we see that the D7-$\overline{{\rm D7}}$ embedding is of the "U-shape" kind that gives the geometric realization of the $U_L(N_f)\times U_R(N_f)$ $\rightarrow$ $U(N_f)$ chiral symmetry breaking. The latter is exactly what will translate into EWSB once an appropriate embedding of the EW $SU(2)\times U(1)$ into $U_L(N_f)\times U_R(N_f)$ is chosen.

Now let us look at the UV region. The metric (\ref{UVM}) implies that:
\be
\det (g_8) = - \frac{2}{3} A^8 H^4 \, e^{16 \rho / 3} \left( \frac{8}{3} + \theta^2_{\rho} + \sin^2 \theta \varphi^2_{\rho} \right) \, ,
\ee
where $H = \frac{2^{-1/3} c \sin \alpha}{4}$. The resulting Lagrangian is:\footnote{Recall that the dilaton is constant.}
\be
{\cal L} = const \times e^{8 \rho / 3} \left( \frac{8}{3} + \theta_{\rho}^2 + \sin^2 \theta \varphi_{\rho}^2 \right)^{1/2} \, .
\ee 
Again, any of $\theta = 0\,, \pm \pi / 2\,, \pi$ is a solution of the $\theta$ equation of motion. Choosing as before $\theta = \pi / 2$, we find that the $\varphi$ equation of motion is:
\be
\varphi_{\rho}^3 + \frac{8}{3} \varphi_{\rho} + \varphi_{\rho \rho} = 0 \, .
\ee
The last equation is solved by:
\be \label{UVsol}
\tan \left( \frac{ 4 \varphi (\rho)}{\sqrt{6}} - C_2 \right) = \pm \left( C_1 e^{16 \rho / 3} - 1 \right)^{1/2} \, ,
\ee
where $C_1$ and $C_2$ are integration constants. 

In moving from the UV to the lower-energy intermediate region, the shape of the solution changes from (\ref{UVsol}) to (\ref{phiEq}). In principle, we can write $C_1$ as $C_1^{-1} = e^{16 \rho_* /3}$, where the integration constant $\rho_*$ is some radial value below the lower end of the range of validity of the metric (\ref{UVM}), so that in the whole UV region there are two separate branches, one for D7 and the other for $\overline{\rm D7}$. Note, however, that both constants $C_{1,2}$ will drop out of the computation of the S-parameter, as we will see in the following.

\subsection{S-parameter}

According to Section 2, in order to compute the S-parameter we first need to calculate the coefficient functions $a$ and $b$ in (\ref{GFAction}). For the intermediate region, using the embedding (\ref{phiEq}), we find that these functions are:
\be
a(\rho) = 2 A^2 \hat{C} C^2 B \, e^{4 \rho} \left( 1 + \frac{e^{4 \rho_0}}{e^{4 \rho} - e^{4 \rho_0}} \right)^{1/2} \, ,
\ee
\be
b(\rho) = \frac{1}{2} A^2 \hat{C} C \left(1 + \frac{e^{4 \rho_0}}{e^{4 \rho} - e^{4 \rho_0} } \right)^{\!-1/2} \, ,
\ee
where $\hat{C} = e^{-\phi}$ and we have used that from (\ref{phiEq}) one has:\footnote{The $\pm$ in (\ref{dvp}) corresponds to the $\pm$ in (\ref{phiEq}). Note also that the expressions, that determine $a(\rho)$ and $b(\rho)$, depend on $\varphi_{\rho}$ only via $\varphi_{\rho}^2$ and so are independent of the sign of $\varphi_{\rho}$.}
\be \label{dvp}
\frac{d \varphi}{d \rho} = \pm 2 \sqrt{B} e^{2 \rho_0} \frac{1}{\sqrt{1 - \frac{e^{4 \rho_0}}{e^{4 \rho}}}}
\ee
and therefore $(4 B e^{4 \rho} + \theta_{\rho}^2 + \sin^2 \theta \varphi_{\rho}^2) = 4 B e^{4 \rho} ( 1 + \frac{e^{4 \rho_0}}{e^{4 \rho} - e^{4 \rho_0}})$ for $\theta = \pi/2$. On the other hand, the embedding solution (\ref{UVsol}) implies:
\be
\frac{d \varphi}{d \rho} = \pm \sqrt{\frac{8}{3}} \,\frac{1}{\sqrt{C_1 e^{16 \rho / 3} -1}} \, .
\ee
Using this and the corresponding induced metric, one can compute that in the UV region:
\be \label{aUV}
a (\rho) = \frac{4}{3} \hat{C} A^2 H^2 e^{8 \rho /3} \left( 1 + \frac{1}{C_1 e^{16 \rho /3} - 1} \right)^{1/2} \, ,
\ee
\be \label{bUV}
b (\rho) = \frac{1}{2} \hat{C} A^2 H e^{4 \rho / 3} \left( 1 + \frac{1}{C_1 e^{16 \rho /3} - 1} \right)^{\!-1/2} \, .
\ee

Now, in order to evaluate (\ref{Sexact}), we need the solution of (\ref{NNMode}) in the UV region. In this region, upon substituting (\ref{aUV}) and (\ref{bUV}), the field equation (\ref{NNMode}) acquires the form:
\be \label{psiUV}
\frac{\pd^2 }{\pd \rho^2} \,\psi^0_{V,A} + \frac{4}{3} \frac{(C_1 e^{16\rho / 3} +1)}{(C_1 e^{16 \rho /3} -1)} \, \frac{\pd }{\pd \rho} \,\psi^0_{V,A} + q^2 \frac{ P \, C_1 e^{20 \rho /3}}{C_1 e^{16 \rho /3} -1} \,\psi^0_{V,A} = 0 \, ,
\ee
where $P = \frac{8}{3} H$. This equation cannot be solved analytically for every $\rho$. However, for $\rho \rightarrow \infty$ things simplify considerably. Namely, since for large enough $\rho$ we have that $C_1 e^{16 \rho /3} >\!\!> 1$ regardless of the value of the still undetermined constant $C_1$, we find that (\ref{psiUV}) reduces to:
\be \label{FEs}
\frac{\pd^2 }{\pd \rho^2} \,\psi^0_{V,A} + \frac{4}{3} \,\frac{\pd }{\pd \rho} \,\psi^0_{V,A} + q^2 P \, e^{4 \rho /3} \,\psi^0_{V,A} = 0 \, .
\ee
The last equation can be solved analytically and the solution is:
\be \label{psiSol}
\psi^0_{V,A} (q^2, \rho) = const_1 \,e^{-2 \rho /3} \,J_1 \!\!\left( \frac{3}{2} q \sqrt{P} e^{2\rho /3} \right) + \,const_2 \,e^{-2\rho /3} \,Y_1 \!\!\left( \frac{3}{2} q \sqrt{P} e^{2\rho /3} \right) \,,
\ee
where $J$ and $Y$ are the Bessel functions of the first and second kind respectively. Note that, as in Section 3, the constants of integration can actually depend on $q$, since the latter is just a parameter in the differential equation (\ref{FEs}). This point will be of significance later on.

\subsubsection{V vs A modes}

Before analyzing the implications of the above solution, let us first comment on the issue of V vs A modes. Recall that $\rho$ is the background radial variable and so, when $\rho$ runs in the interval $(\rho_0 , \infty)$, it does not distinguish between the two branches of the D7-$\overline{{\rm D}7}$ embedding. To distinguish between them, one needs to change variables from $\rho$ to a worldvolume coordinate $z$ that runs only over the "U-shape" embedding, i.e. $z \in (-\infty, \infty)$ with $z$ positive being the brane and $z$ negative being the antibrane branch. In terms of such a variable, the vector modes are defined as the ones that are symmetric under $z \rightarrow -z$, whereas the axial-vector modes are those that are antisymmetric under this transformation. Let us define this worldvolume variable as:
\be \label{ChV}
z^2 = \rho^2 - \rho_0^2 \, .
\ee 
To see how the equation of motion for $\psi_{V,A}^0 (q^2,z)$ looks like, let us perform a change of variables $\rho \rightarrow \rho(z)$ in the action (\ref{GFAction}). We see that 
\be
\int d \rho \left[ a(\rho) F_{\mu \nu} F^{\mu \nu} + 2 b(\rho) F_{\mu \rho} F^{\mu}{}_{\rho}\right] = \int d z \left[ \hat{a}(z) F_{\mu \nu} F^{\mu \nu} + 2 \hat{b}(z) F_{\mu z} F^{\mu}{}_{z}\right] \, ,
\ee
where 
\be
\hat{a}(z) = a(\rho) \,\pd_z \rho \qquad {\rm and} \qquad \hat{b}(z) = \frac{b(\rho)}{\pd_z \rho} \,\, .
\ee
Therefore, the field equation is:
\be
\frac{1}{\hat{a}(z)} \, \pd_z \!\!\left[ \hat{b}(z) \,\pd_z \psi^0_{V,A} (q^2,z) \right] = -q^2 \psi^0_{V,A} (q^2,z) \, .
\ee
For the change of variables (\ref{ChV}) and in the large $z$ limit this gives:
\be
\frac{\pd^2 }{\pd z^2} \,\psi_{V,A}^0 + \frac{4}{3} \frac{z}{|z|} \,\frac{\pd}{\pd z} \,\psi_{V,A}^0 + q^2 P e^{4 |z|/3} \psi_{V,A}^0 = 0 \, .
\ee
Clearly, for $z>0$ (the D7 branch) this equation is exactly the same as (\ref{FEs}). On the other hand, for $z<0$ (the $\overline{{\rm D}7}$ branch) one has:
\be
\frac{\pd^2 }{\pd z^2} \,\psi_{V,A}^0 - \frac{4}{3} \,\frac{\pd}{\pd z} \,\psi_{V,A}^0 + q^2 P e^{- 4 z/3} \psi_{V,A}^0 = 0 \, .
\ee
Hence, at large negative $z$ the symmetric solutions have the form
\be \label{zn}
\psi_V^0 (q^2,z)|_{z<0} = \hat{C}_1^V \,e^{2 z /3} \,J_1 \!\!\left( \frac{3}{2} q \sqrt{P} e^{-2 z /3} \right) + \,\hat{C}_2^V \,e^{2 z /3} \,Y_1 \!\!\left( \frac{3}{2} q \sqrt{P} e^{-2 z /3} \right) ,
\ee
where the constants $\hat{C}_1^V$, $\hat{C}_2^V$ are the same as those in the large positive $z$ limit, in which
\be \label{zp}
\psi_V^0 (q^2,z)|_{z>0} = \hat{C}_1^V \,e^{-2 z /3} \,J_1 \!\!\left( \frac{3}{2} q \sqrt{P} e^{2 z /3} \right) + \,\hat{C}_2^V \,e^{-2 z /3} \,Y_1 \!\!\left( \frac{3}{2} q \sqrt{P} e^{2 z /3} \right) .
\ee
Similarly, at large $z$ the antisymmetric modes $\psi_A^0 (q^2,z)$ have the same form as (\ref{zn}) and (\ref{zp}), with the only difference that if we denote the constants entering the positive $z$ asymptotics by $\hat{C}_1^A$ and $\hat{C}_2^A$, then those appearing in the negative $z$ asymptotics are $-\hat{C}_1^A$ and $-\hat{C}_2^A$ respectively. 

The introduction of the variable $z$ is crucial for the study of the solutions around $z \approx 0$, in order to impose the conditions $\psi_A^0 (q^2,0) = 0$ and $\pd_z \psi_V^0 (q^2,z) |_{z=0} = 0$ that define the symmetric and antisymmetric modes respectively. However, for the present considerations of the large distance region it is not significant, as the derivation of the formula (\ref{Sexact}) already used that the contribution of both branches is equal to twice the contribution of just one of them. In particular, the modes in (\ref{Sexact}) are only those on one of the two branches. So in the present section we will concentrate only on one (the D7) branch and will continue using the variable $\rho$, in order not to deal with more cumbersome expressions involving $|z|$.

\subsubsection{Small $q^2$ solution} \label{SmallqSol}

Let us now get back to the computation of the S-parameter. To understand the behaviour of the solution (\ref{psiSol}) in the limits of interest for us, let us first recall the asymptotic behaviour of the Bessel functions at large and at small argument. For large argument (i.e., $y  >\!\!> 1$), one has:
\be \label{LargeArg}
J_{\alpha} (y) \approx \sqrt{\frac{2 \pi}{y}} \cos \left( y - \frac{\alpha \pi}{2} - \frac{\pi}{4} \right) \, , \qquad Y_{\alpha} (y) \approx \sqrt{\frac{2 \pi}{y}} \sin \left( y - \frac{\alpha \pi}{2} - \frac{\pi}{4} \right) \, ,
\ee 
whereas for small argument (i.e., $y <\!\!< 1$) the asymptotic behaviour is:
\be \label{SmallArg}
J_1 (y) = \frac{1}{2} y - \frac{1}{16} y^3 + O(y^5) \, , \quad Y_1 (y) = -\frac{2}{\pi} \frac{1}{y} + \frac{2 \ln (y/2) +2 \gamma - 1}{\pi} y + O(y^3 \ln y) \, .
\ee

Now, at first sight it is not clear which of this two limits, if any, is relevant for us, since we are interested in taking {\it both} $q \rightarrow 0$ {\it and} $\rho \rightarrow \infty$, in which case the argument of the Bessel functions in (\ref{psiSol}) is of the form $0 \times \infty$. To resolve this ambiguity, recall that we are looking for the small-$q^2$ behaviour of the {\it non}-normalizable modes. Whether a function $\psi (\rho)$ is normalizable or not is determined by whether $\int_0^{\infty} d\rho \,g_{\rho \rho} \,\psi^2$ is finite or not. Reading off the UV behaviour of $g_{\rho \rho}$ from the metric (\ref{UVM}), we see that for finite $q$ functions of the form (\ref{LargeArg}) are normalizable. To understand the significance of this, recall that the normalizable modes satisfy the same field equation as the non-normalizable ones, only with the substitution $q^2 \rightarrow m^2$; see (\ref{normFE}), (\ref{NNMode}). Hence the solution for them is also (\ref{psiSol}), where $q$ is substituted by the finite mass parameter $m$. The discrete spectrum $m_n$, $n=1,2,...$ arises because only for discrete values one can match the asymptotic solutions with those that satisfy the appropriate boundary conditions around $\rho_0$ (i.e., are either symmetric or antisymmetric w.r.t. reflection around $\rho_0$ on the worldvolume of the "U-shaped" D7-$\overline{{\rm D7}}$). Hence, we have identified (\ref{LargeArg}) as the asymptotic behaviour of the normalizable modes. This leaves us with (\ref{SmallArg}) as the candidate for the large distance asymptotics of the non-normalizable modes for small $q$; we will confirm this below in a different manner.

In the above paragraph we have treated $q$ as finite, whereas $\rho$ was taken to be infinitly large. This is justified because normalizability is a generic question about the behaviour at large distance. However, now we want to turn to the small-$q$ behaviour of the non-normalizable modes. So let us introduce an upper limit (a UV cut-off from the field theory perspective) for $\rho$, i.e. take $\rho \in (0, \rho_{\Lambda})$ with $\rho_{\Lambda}$ large but finite, and treat $q$ as infinitesimal. Of course, at the end, the S-parameter should not depend on the value of $\rho_{\Lambda}$. Now, in the limit $q \rightarrow 0$ and $\rho \rightarrow \rho_{\Lambda}$ the argument of the Bessel functions in (\ref{psiSol}) is small and we have the expansions (\ref{SmallArg}). In principle, plugging those expressions in (\ref{Sexact}) should give the value of the S-parameter. In practice however, just as in the regular techinicolor example of Section 3, one obtains either zero or infinity, depending on whether one takes only one or both Bessel functions in (\ref{psiSol}). As we saw in the previous section, this problem can be overcome by regulating the solution (\ref{psiSol}) in a manner that results in a solution at small $q$, which tends to a constant at large distance, rather than to $0$ or $\infty$. As in \cite{HolTech}, we will take this constant to be 1. 

So we want to find a solution of (\ref{FEs}), which tends to 1 and is an expansion in powers of small $q^2$ with the variable $\rho$ satisfying $\rho \le \rho_{\Lambda}$. Since we are looking for a solution of (\ref{FEs}) valid at small $q$, let us first consider the equation
\be \label{HomFE}
\frac{\pd^2 }{\pd \rho^2} \,\psi^0 (q^2,\rho) + \frac{4}{3} \,\frac{\pd }{\pd \rho} \,\psi^0 (q^2,\rho) = 0 \,.
\ee
Its most general solution is $f_1 (q^2) + f_2 (q^2) e^{-4 \rho/3}$, where $f_1 (q^2)$ and $f_2 (q^2)$ are independent of $\rho$ but can be any functions of $q$. Since we want a solution that is an expansion in small $q^2$ and tends to $1$, this freedom is reduced for us to the following:
\be \label{psiHom}
\psi^0_h = 1 + \tilde{C}_1 q^2 + (\tilde{C}_2 + \tilde{C}_3 q^2) e^{-4 \rho /3} + {\cal O} (q^4) \, ,
\ee 
where $\tilde{C}_1$, $\tilde{C}_2$ and $\tilde{C}_3$ are constants and, as before, we have stopped at ${\cal O} (q^2)$ since the higher order terms do not contribute in the expression for the S-parameter (\ref{Sexact}). Clearly, we can solve (\ref{FEs}) to order $q^2$ by adding to $\psi_h^0$ a particular solution of the following inhomogeneous equation:
\be \label{InhFE}
\frac{\pd^2 }{\pd \rho^2} \,\psi^0 (q^2,\rho) + \frac{4}{3} \,\frac{\pd }{\pd \rho} \,\psi^0 (q^2,\rho) + q^2 P e^{4 \rho /3} + q^2 P \tilde{C}_2 = 0 \, ,
\ee
where the inhomogeneous terms are the zeroth order contribution in the $q^2$ expansion  of $\psi_h^0$ in (\ref{psiHom}) multiplied by the coefficient of the last term in (\ref{FEs}). It is easily seen that a particular solution of the inhomogeneous equation (\ref{InhFE}) is given by
\be
\psi_i^0 = - \frac{9}{32} q^2 P e^{4\rho /3} - \frac{3}{4} \tilde{C}_2 q^2 P \rho \, .
\ee
Hence, adding $\psi_h^0$ and $\psi_i^0$, we obtain:
\be \label{Solqexp}
\psi^0_{V,A} = 1 + \tilde{C}_1^{V,A} q^2 + (\tilde{C}_2^{V,A} + \tilde{C}_3^{V,A} q^2) e^{-4 \rho /3} - \frac{9}{32} q^2 P e^{4\rho /3} - \frac{3}{4} \tilde{C}_2^{V,A} q^2 P \rho + {\cal O} (q^4) \, .
\ee
Substituting these solutions in the formula for the S-parameter, we find that:
\bea \label{Sfinal}
\hspace*{-1cm}S \!\!&=& \!\!- 8 \pi \kappa \left[ b(\rho) \frac{\pd}{\pd q^2} \left( \psi_V^0 \pd_{\rho} \psi_V^0 - \psi_A^0 \pd_{\rho} \psi_A^0 \right) \right]_{\rho = \rho_{\Lambda}, \,q^2 = 0} \nn \\
\!\!&=& \!\!- 8 \pi \kappa \,\hat{B} \left[ -\frac{4}{3} ( \tilde{C}_3^V + \tilde{C}_1^V \tilde{C}_2^V - \tilde{C}_3^A - \tilde{C}_1^A \tilde{C}_2^A ) - \frac{3}{4} P ( \tilde{C}_2^{V \,2} - \tilde{C}_2^{A \,2} ) \right. \nn \\ 
\!\!&-& \!\!\left. \frac{3}{4} P ( \tilde{C}_2^V - \tilde{C}_2^A ) e^{4 \rho_{\Lambda} /3} + P  ( \tilde{C}_2^{V \,2} - \tilde{C}_2^{A \,2} ) \rho_{\Lambda} - \frac{8}{3} (\tilde{C}_2^V \tilde{C}_3^V - \tilde{C}_2^A \tilde{C}_3^A ) e^{-4 \rho_{\Lambda} /3} \right] \!,
\eea
where $\hat{B} = \frac{1}{2} \hat{C} A^2 H$ is the constant coefficient of $b(\rho)$ in (\ref{bUV}). 

Clearly, the above answer for the S-parameter diverges in the limit $\rho_{\Lambda} \rightarrow \infty$, unless either $\tilde{C}_2^V = \tilde{C}_2^A$ or both $\tilde{C}_2^{V,A}$ vanish. However, neither of those two options is possible for the following reason. Recall from (\ref{PiVA}) that: 
\bea 
\Pi_V (q^2) \!\!&=& \!\!2\kappa \left[ b(\rho) \,\psi_V^0 (q^2, \rho) \,\pd_{\rho} \psi_V^0 (q^2, \rho) \right]_{\rho = \infty} \,\, , \nn \\
\Pi_A (q^2) \!\!&=& \!\!2\kappa \left[ b(\rho) \,\psi_A^0 (q^2, \rho) \,\pd_{\rho} \psi_A^0 (q^2, \rho) \right]_{\rho = \infty} \,\, .
\eea
Using (\ref{Solqexp}), this implies that $\Pi_V (q^2 = 0) = const \times \tilde{C}_2^V$ and $\Pi_A (q^2 = 0) = const' \times \tilde{C}_2^A$. On the other hand, as is well-known, $\Pi_A (0) = F_{\pi}^2$ with $F_{\pi} = 250 \,{\rm GeV}$ being the technipion decay constant, whereas $\Pi_V (0) = 0$. Hence we must always have $\tilde{C}_2^A \neq \tilde{C}_2^V = 0$.\footnote{Despite that, we will keep writing $\tilde{C}_2^V$ in the following, in order to maintain explicit symmetry between V and A modes.}

Therefore, to obtain a finite answer for the S-parameter, we need to renormalize the gravity action. We will do that in Section \ref{Ren}. At this point, let us make the following interesting observation. The leading divergence, arising from the $V$ and $A$ contributions, is in fact $\sim e^{8 \rho_{\Lambda}/3}$. However, just as in the example of Section 3, it is the same for both $V$ and $A$ modes and thus cancels in the difference. Another important observation follows from understanding the relation between the small $q^2$ solution (\ref{Solqexp}) and the general solution for any $q$ in (\ref{psiSol}). So let us now turn to that issue.

\subsubsection{Relation to general solution} \label{RelGenSol}

We argued earlier that the small $q$ behaviour of the terms in the general solution (\ref{psiSol}) should be given by (\ref{SmallArg}).\footnote{We are again keeping in mind the cut-off $\rho_{\Lambda}$. So we view $\rho$ as finite, no matter how large, and $q$ as infinitesimal.} According to those expansions, in the $q \rightarrow 0$ limit one has $J_1 (\frac{3}{2}q \sqrt{P} e^{2\rho/3}) \rightarrow 0$ and $Y_1 (\frac{3}{2} q \sqrt{P} e^{2 \rho/3}) \rightarrow \infty$. We will regulate these solutions in the same manner as in Section 3.

More precisely, we will consider the following rescaling:
\be \label{rescale}
e^{-2 \rho /3} J_1 \!\left( \frac{3}{2} q \sqrt{P} e^{2 \rho /3} \right) \rightarrow \,\frac{e^{-2 \rho /3} J_1 \!\left( \frac{3}{2} q \sqrt{P} e^{2 \rho /3} \right)}{e^{-2 \rho_{\bullet} /3} J_1 \!\left( \frac{3}{2} q \sqrt{P} e^{2 \rho_{\bullet} /3} \right)} \,\, ,
\ee
where $\rho_{\bullet}$ is some finite fixed value completely unrelated to the cut-off $\rho_{\Lambda}$. Using the expansion (\ref{SmallArg}) and expanding the whole ratio in (\ref{rescale}) in small $q$, we find:
\be \label{RegExpJ}
\frac{e^{-2 \rho /3} J_1 \!\left( \frac{3}{2} q \sqrt{P} e^{2 \rho /3} \right)}{e^{-2 \rho_{\bullet} /3} J_1 \!\left( \frac{3}{2} q \sqrt{P} e^{2 \rho_{\bullet} /3} \right)} = 1 + q^2 \frac{9 P}{32} e^{4 \rho_{\bullet} /3} - q^2 \frac{9 P}{32} e^{4 \rho/3} + {\cal O} (q^4) \, .
\ee
Clearly, each of the terms in (\ref{RegExpJ}) corresponds to a term in (\ref{Solqexp}). However, the latter has additional terms that did not appear in the regulated $J_1$ solution. It is easy to see that those come from regulating the $Y_1$ term in (\ref{psiSol}). Indeed, we find:
\be \label{RegExpY}
\frac{e^{-2 \rho /3} Y_1 \!\left( \frac{3}{2} q \sqrt{P} e^{2 \rho /3} \right)}{e^{-2 \rho_{\bullet} /3} Y_1 \!\left( \frac{3}{2} q \sqrt{P} e^{2 \rho_{\bullet} /3} \right)} = e^{4 \rho_{\bullet} /3} e^{- 4 \rho /3} - \frac{3}{4} q^2 P e^{4 \rho_{\bullet} /3} \!\left( C_Y^1 + C_Y^2 e^{- 4 \rho/3} \right) - \frac{3}{4} q^2 P e^{4 \rho_{\bullet} /3} \rho \!+ {\cal O} (q^4)\, ,
\ee
where
\bea
C_Y^1 &=& \frac{3}{4} (2 \gamma -1) + \frac{3}{2} \ln \left( \frac{3 q \sqrt{P}}{4} \right) \, , \nn \\
C_Y^2 &=& e^{4 \rho_{\bullet} /3} \left[ \frac{3}{4} (2 \gamma -1) + \frac{3}{2} \ln \left( \frac{3 q \sqrt{P} e^{2 \rho_{\bullet} /3}}{4} \right) \right] \, .
\eea
Note however, that unlike (\ref{Solqexp}), the expression (\ref{RegExpY}) contains terms of the form $q^2 \ln q$. The latter can be canceled by additional multiplication by $q$-dependent expressions that are constant w.r.t. to $\rho$. Namely, multiplying (\ref{RegExpY}) by $(1+\frac{9}{8} P e^{4 \rho_{\bullet} /3} q^2 \ln q)$ cancels the $q^2 \ln q$ term in $C_Y^2$, without affecting the rest of the expansion to order $q^2$. Similarly, in order to cancel the $q^2 \ln q$ term in $C_Y^1$, we multiply (\ref{RegExpJ}) by $(1+\frac{9}{8} P e^{4 \rho_{\bullet} /3} q^2 \ln q)$.

As in Section 3, note that, in general, we can have different values of $\rho_{\bullet}$ for the $J_1$ and $Y_1$ solutions. I.e. $\rho_{\bullet}^J \neq \rho_{\bullet}^Y$, which is rather important for the matching of the number of constants in the small $q$ solution (\ref{Solqexp}) and the small $q$ expansion of the general solution $\hat{C}_1 e^{-2 \rho /3} J_1 + \hat{C}_2 e^{-2 \rho /3} Y_1$. Since the constant piece in (\ref{Solqexp}) is normalized to $1$, then (\ref{RegExpJ}) implies that $\hat{C}_1 = 1$. Hence the three independent constants, corresponding to $\tilde{C}_{1,2,3}$, are  $\hat{C}_2$, $\rho_{\bullet}^J$ and $\rho_{\bullet}^Y$. To recapitulate, the solution
\bea
\psi^0 (q^2, \rho ) &=& \frac{\left( 1+\frac{9}{8} \hat{C}_2 P e^{\frac{4 \rho^Y_{\bullet}}{3}} \,q^2 \ln q \right) }{e^{-\frac{2 \rho^J_{\bullet}}{3}} \,J_1 \!\left( \frac{3}{2} q \sqrt{P} e^{\frac{2 \rho^J_{\bullet}}{3}} \right)} \,\,\, e^{-2 \rho /3} \,\, J_1 \!\left( \frac{3}{2} q \sqrt{P} e^{2\rho /3} \right) \nn \\ \nn \\
&+& \frac{ \hat{C}_2 \left( 1+\frac{9}{8} P e^{\frac{4 \rho^Y_{\bullet}}{3}} \,q^2 \ln q \right) }{e^{-\frac{2 \rho^Y_{\bullet}}{3}} \,\, Y_1 \!\left( \frac{3}{2} q \sqrt{P} e^{\frac{2 \rho^Y_{\bullet}}{3}} \right)} \,\,\, e^{-2\rho /3} \,\,\, Y_1 \!\left( \frac{3}{2} q \sqrt{P} e^{2\rho /3} \right)
\eea
has small $q$ expansion exactly of the form (\ref{Solqexp}) with 
\bea \label{Ct}
\tilde{C}_1 &=& \frac{9}{32} P e^{4 \rho^J_{\bullet} /3} - \frac{3}{4} P e^{4 \rho^Y_{\bullet} \!/3} \hat{C}_2 \left[ \frac{3}{4} (2 \gamma - 1) + \frac{3}{2} \ln \left( \frac{3 \sqrt{P}}{4} \right) \right] \, , \nn \\
\tilde{C}_2 &=& e^{4 \rho^Y_{\bullet} \!/3} \hat{C}_2 \,\, , \nn \\
\tilde{C}_3 &=& - \frac{3}{4} P e^{8 \rho^Y_{\bullet} \!/3} \hat{C}_2 \left[ \frac{3}{4} (2 \gamma - 1) + \frac{3}{2} \ln \left( \frac{3 \sqrt{P} e^{2 \rho^Y_{\bullet} \!/3}}{4} \right) \right] \, .
\eea
We have thus understood how (\ref{Solqexp}) arises from the general solution (\ref{psiSol}) in the limit of small $q$ and for a particular choice of the integration constants. 

Notice that (\ref{Ct}) implies that the coefficient $\tilde{C}_1$ receives contributions from both the $J_1$ and the $Y_1$ terms, whereas each of the coefficients $\tilde{C}_{2}$ and $\tilde{C}_3$ comes entirely from the $Y_1$ term. This again underscores the observation we made in Section 3, that the intuition to disregard the diverging in the $q \rightarrow 0$ limit $Y_1$ solution is incorrect. In fact, dropping the $Y_1$ solution would have led to an identical zero as each of the terms in the result (\ref{Sfinal}) is proportional to either $\tilde{C}_2$ or $\tilde{C}_3$. It is also worth noting that each of the constants $\tilde{C}_1$ and $\tilde{C}_3$ is proportional to $P$.

\subsection{Renormalization} \label{Ren}

Holographic renormalization \cite{HolRen} was developed in the context of the AdS/CFT correspondence as an intrinsic way of taming infrared (IR) divergences that occur in gravitational backgrounds. Recall that large distances on the gravity side correspond to high energies on the dual field theory side. So the gravitational IR divergences are the natural counterpart of the field theoretic UV divergences.   

Prior to holographic renormalization, IR divergences were removed from gravitational actions by so called background subtraction \cite{GH}. Namely, by choosing a particular reference background and subtracting the action for this reference space-time from the action for the space-time of interest. However, this procedure is ambiguous (as there may be more then one candidate for a reference background) and not always applicable (as in some cases there is no isometric embedding of the regulating boundary into the reference space-time). On the other hand, holographic renormalization is a well-defined intrinsic procedure, that consists of adding new terms on the regulating boundary of the background of interest. These, so called, counterterms can be deduced from the requirement that the total gravitational action be finite, but they also follow from requiring that the variational principle be well-defined.

Initially, holographic renormalization was developed for a very limited class of gravitational backgrounds, namely asymptotically AdS ones; see \cite{KSk}. However, recent works have extended the classes of backgrounds, to which this method can be applied, in interesting directions. Most relevant for us, asymptotically linear dilaton (ALD) gravity backgrounds, of which the background in our Section \ref{GrBackground} is a special case, can be renormalized by adding appropriate counterterms \cite{MM}. The reason the background of interest for us falls within this class is that ALD backgrounds occur as near-horizon limits of stacks of D5 or NS5 branes in string theory. Recall that the background of \cite{NPP} is a deformation of the original Maldacena-Nunez (MN) solution \cite{MN} and both solutions (i.e. the familiar MN background and the deformated one) arise from stacks of 5-branes wrapping an $S^2$. Let us also note that linear dilaton asymptotics, i.e. $\phi \sim \rho$ at large $\rho$, actually refers to the asymptotic behaviour of the dilaton in a coordinate system (in string frame) that is different from the one used in \cite{MM}.\footnote{For comparison of the two coordinate systems, see for example \cite{MV}.} In the conventions of \cite{MM}, at large $\rho$ in an ALD background the dilaton behaves as $\phi \sim c_1 \log \rho + c_2 + ...$ with $c_{1,2}$ constants. Clearly, an asymptotically constant dilaton (which is the case of interest for us) is a special case of this behaviour for vanishing coefficient of the $\log \rho$ term. 

Without going into details, let us just state here the essence of the result of \cite{MM}. A gravitational theory with an (Einstein frame) action of the form:
\be
I = \int_{{\cal M}} d^{d+1} x \sqrt{g} \left( R - \frac{4}{d-1} \nabla^{\mu} \phi \nabla_{\mu} \phi - \frac{1}{2 p!} e^{2 \alpha \phi} F^{\mu_1 ... \mu_p} F_{\mu_1 ... \mu_p} \right) \, ,
\ee
where $\phi$ is a scalar field and $F$ is a $p$-form, and with ALD asymptotics for the fields\footnote{For details on the required asymptotic behaviour of the fields, that plays the role of boundary conditions, see \cite{MM}} can be renormalized by adding a counterterm of the form:
\be
I_{{\rm CT}} = \int_{\partial {\cal M}} d^d x \sqrt{h} \left[ c_1 e^{- \frac{\alpha}{p-1} \phi} + c_2 e^{\frac{\alpha}{p-1}} \left( R - \frac{1}{2p!} e^{2 \alpha \phi} F^2 \right) \right] \, ,
\ee
where $c_1$ and $c_2$ are constants determined by the finiteness of the total action $I + I_{{\rm CT}}$.\footnote{Alternatively, $I_{{\rm CT}}$ follows from requiring a well-defined variational principle, as mentioned above.} To specialize these general considerations to our case, one needs to take $\alpha=1/2$ and $p=3$ and, further, to identify $\phi$ with the dilaton and $F_3$ with the RR 3-form field strength of type IIB.  

In fact, all we need to take away from the above paragraph is that the background of interest for us is renormalizable and so it is sensible to study probes in it and their implications for the dual field theory. What we actually need to understand in detail is how to renormalize the action of a probe brane in that background. To do that, let us first look at the DBI action of a ${\rm D}7$ probe with no world-volume gauge fields turned on:
\be \label{DBIgdiv}
S_{DBI} = - T_7 \int d^{8} x \sqrt{- {\rm det} g} = - T_7 \Omega_3 \int d^4 x \,d \rho \,\sqrt{- {\rm det} g} \,\, ,
\ee
where we have absorbed the constant (for us) factor $e^{-\phi}$ into the brane tension $T_7$ and, in the second equality, we have integrated over the compact directions wrapped by the D7 brane.
This action can be renormalized by adding a counterterm of the form
\be \label{CTg}
S_{DBI}^{{\rm CT}} = - T_7 \Omega_3 \int d^4 x \sqrt{- {\rm det} \gamma} \times c \, ,
\ee
where $\gamma$ is the metric induced on a regulating surface defined by constant $\rho$, in the notation of our Section \ref{SmallqSol} this surface is given by $\rho = \rho_{\Lambda}$, and $c$ is an appropriately chosen constant that depends on $\rho_{\Lambda}$. The counterterm (\ref{CTg}) has been used recently in \cite{HKKL}. It is easy to see that in our case the large $\rho$ divergence of (\ref{DBIgdiv}) is canceled by taking $c=-\frac{1}{2} A^2 H^2 e^{\frac{8 \rho_{\Lambda}}{3}}$.

Finally, we are ready to turn to the case of interest, namely the DBI action (\ref{YM}):
\be \label{YMdiv}
S_{DBI} = - \frac{\kappa}{4} \int d^4 x \,d \rho \,\sqrt{- {\rm det} g} \,\, g^{ab} g^{cd} F_{ac} F_{bd} \, ,
\ee
where the indices $a,b,c,d$ run over the coordinates $x^{\mu}$ and $\rho$. In view of (\ref{CTg}), it is natural to expect that the action (\ref{YMdiv}) can be renormalized by the addition of a counterterm of the form:
\be \label{CTa}
S_{DBI}^{{\rm CT}} = - \,c \,\frac{\kappa}{4} \int d^4 x \sqrt{- {\rm det} \gamma} \,\, \gamma^{\mu \nu} \gamma^{\mu' \nu'} F_{\mu \mu'} F_{\nu \nu'} \, ,
\ee
where $c$ is an appropriately chosen constant that depends on $\rho_{\Lambda}$. Indeed, we will show now that this is exactly what happens. Using the decomposition (\ref{decomp}) for $\rho = \rho_{\Lambda}$ and the boundary conditions $\psi_{V_n,A_n} (\rho_{\Lambda}) = 0$ and repeating the same kind of considerations as in Section \ref{HolTechSec}, it is easy to see that the counterterm (\ref{CTa}) gives:
\bea \label{CTf}
S_{DBI}^{{\rm CT}} &=& - \,c \,\frac{\kappa}{4} \int d^4 x \left\{ |F_{\mu \nu}^{{\cal V}} (q)|^2 \!\left( \psi^0_V (\rho_{\Lambda}) \right)^2 + |F_{\mu \nu}^{{\cal A}} (q)|^2 \!\left( \psi^0_A (\rho_{\Lambda}) \right)^2 \right\} \nn \\
&=& - \,c \,\frac{\kappa}{4} \int d^4 x \left\{ 2 q^2 |{\cal V}_{\mu}|^2 \!\left( \psi^0_V (\rho_{\Lambda}) \right)^2 + 2 q^2 |{\cal A}_{\mu}|^2 \!\left( \psi^0_A (\rho_{\Lambda}) \right)^2 \right\} \, .
\eea
Now, the renormalized probe-brane action should give finite S-parameter. Since the latter is what we are really interested in and, furthermore, we have already computed its divergences explicitly (see (\ref{Sfinal})), let us fix the constant $c$ directly at the level of the S-parameter. The renormalized S-parameter is obtained from
\be
\Pi_V^{ren} (q^2) = - \frac{\delta}{\delta {\cal V}_{\mu}} \frac{\delta}{\delta {\cal V}_{\nu}} \left( S_{DBI} + S_{DBI}^{{\rm CT}} \right) \Big|_{{\cal V}=0}
\ee
and similarly for $\Pi_A^{ren}$. So the contribution to the S-parameter that is due to (\ref{CTf}) is:
\be \label{SCT}
S^{{\rm CT}} = 8 \pi \kappa \,c \left[ (\psi_V^0)^2 - (\psi_A^0)^2 \right]_{\rho = \rho_{\Lambda}, q^2 = 0} \, ,
\ee
where the numerical coefficient is 8, instead of 4, because of the two branches ${\rm D}7$ and $\overline{{\rm D}7}$. Now, using the solutions (\ref{Solqexp}), we can see that:
\be
\left[ (\psi_V^0)^2 - (\psi_A^0)^2 \right]_{\rho = \rho_{\Lambda}, q^2 = 0} = \left( \tilde{C}_2^{V\,2} - \tilde{C}_2^{A\,2} \right) e^{-\frac{8 \rho_{\Lambda}}{3}} + 2 \left( \tilde{C}_2^V - \tilde{C}_2^A \right) e^{-\frac{4 \rho_{\Lambda}}{3}} \, .
\ee
Hence the $e^{\frac{4 \rho_{\Lambda}}{3}}$ divergence in (\ref{Sfinal}) can be canceled by a counterterm of the form:
\be \label{Div1}
c_1 e^{\frac{8 \rho_{\Lambda}}{3}} \left[ (\psi_V^0)^2 - (\psi_A^0)^2 \right]_{\rho = \rho_{\Lambda}, q^2 = 0} = c_1 \left( \tilde{C}_2^{V\,2} - \tilde{C}_2^{A\,2} \right) + 2 c_1 \left( \tilde{C}_2^V - \tilde{C}_2^A \right) e^{\frac{4 \rho_{\Lambda}}{3}} \, ,
\ee
where $c_1 = - \frac{3}{8} P \hat{B}$. On the other hand, the $\rho_{\Lambda}$ divergence in (\ref{Sfinal}) can be canceled by
\be \label{Div2}
c_2 \rho_{\Lambda} e^{\frac{4 \rho_{\Lambda}}{3}} \left[ (\psi_V^0)^2 - (\psi_A^0)^2 \right]_{\rho = \rho_{\Lambda}, q^2 = 0} = c_2 \left( \tilde{C}_2^{V\,2} - \tilde{C}_2^{A\,2} \right) \rho_{\Lambda} \,e^{-\frac{4 \rho_{\Lambda}}{3}} + 2 c_2 \left( \tilde{C}_2^V - \tilde{C}_2^A \right) \rho_{\Lambda}
\ee
with $c_2 = \frac{1}{2} ( \tilde{C}_2^V + \tilde{C}_2^A ) P \hat{B}$. From (\ref{Div1}) and (\ref{Div2}), we conclude that to cancel all divergences we have to take the constant $c$ in (\ref{SCT}) to be:
\be \label{cfixed}
c = - \hat{B} P \left( \frac{3}{8} e^{\frac{8 \rho_{\Lambda}}{3}} - \frac{1}{2} ( \tilde{C}_2^V + \tilde{C}_2^A ) \rho_{\Lambda} \,e^{\frac{4 \rho_{\Lambda}}{3}} \right) .
\ee 
Note that, not surprisingly, the leading term in $c$ is of the same form as the coefficient needed to renormalize the action (\ref{DBIgdiv}).

To recapitulate, the renormalized S-parameter is obtained by adding (\ref{Sfinal}) and (\ref{SCT}), with $c$ given by (\ref{cfixed}), and taking the limit $\rho_{\Lambda} \rightarrow \infty$. Note that, due to the first term on the right hand side of (\ref{Div1}), the counterterm $S^{{\rm CT}}$ adds a finite contribution to the final answer:
\be
S^{{\rm CT}}_{finite} = - 8 \pi \kappa \hat{B} P \frac{3}{8} \left( \tilde{C}_2^{V\,2} - \tilde{C}_2^{A\,2} \right) \, .
\ee
Hence the renormalized S-parameter is:
\bea \label{Sren}
S_{ren} &=& \left( S + S^{CT} \right)\!\big|_{\rho_{\Lambda} \rightarrow \infty} \nn \\
&=& - 8 \pi \kappa \,\hat{B} \left[ -\frac{4}{3} ( \tilde{C}_3^V + \tilde{C}_1^V \tilde{C}_2^V - \tilde{C}_3^A - \tilde{C}_1^A \tilde{C}_2^A ) - \frac{3}{8} P ( \tilde{C}_2^{V \,2} - \tilde{C}_2^{A \,2} ) \right] .
\eea
So far, we kept $\tilde{C}_2^V$ explicitly in order to maintain symmetry between V and A modes. However, as already pointed out at the end of Section \ref{SmallqSol}, we should actually set $\tilde{C}_2^V = 0$. So, the final result is:
\be
S_{ren} = - 8 \pi \kappa \,\hat{B} \left[ -\frac{4}{3} ( \tilde{C}_3^V - \tilde{C}_3^A - \tilde{C}_1^A \tilde{C}_2^A ) + \frac{3}{8} P ( \tilde{C}_2^A )^2 \right] .
\ee
Note, again, that we would have missed the term $\tilde{C}_1^A \tilde{C}_2^A$, had we substituted the boundary condition $\psi^0 = 1$ in (\ref{Sexact}) before taking the limit $\rho \rightarrow \infty$.

\section{Discussion} \label{Disc}

We studied a model of walking technicolor, obtained by embedding ${\rm D}7$-$\overline{{\rm D}7}$ probes in the background of \cite{NPP}. We were able to show that one can extract a finite answer for the S-parameter by using holographic renormalization. However, we could not determine analytically the numerical constants $\tilde{C}_{1,3}$ above. Since the latter will depend on the parameters of the gravity background, at this stage it is premature to make conclusions about the value of $S_{ren}$ or its dependence on the length of the walking region. Calculating numerically $\tilde{C}_{1,3}$ and exploring their dependence on background parameters is work in progress \cite{ASW}. It is worth noting that \cite{ModWalk} considered a type IIB background, which is a modification of the one in \cite{NPP} with a similar walking region but a different UV one. It would be interesting to apply the methods we used here, in order to investigate that modified background and to see whether this would produce a finite (after renormalization) S-parameter. If yes, then it could be instructive to explore the differences and similarities with the case studied here. 

Finally, here we have only concentrated on the technicolor sector. However, in order to obtain a complete picture, one would have to include the Standard Model fields, presumably via other probe branes embedded in the same background. It would be very interesting to explore this and related issues. In particular, one such issue is the contribution of extended technicolor \cite{ETC} gauge bosons to the S-parameter. It was argued in \cite{KSY}, that the latter would be a rather small effect. It would, clearly, be interesting to reproduce that from the gravity side. Also, it would be worth verifying with our methods the lower bound for the S-parameter, that was suggested in \cite{FS} based on purely field theoretic arguments.

\section*{Acknowledgements}

I would like to thank L. C. R. Wijewardhana for many illuminating discussions and for reading the draft. I am also grateful to B. Acharya, P. Argyres, A. Buchel, M. Kruczenski, D. Minic, A. Parnachev, R. Shrock and T. Takeuchi for useful conversations and J. Erlich, O. Mintakevich, M. Piai and C. Nunez for correspondence. In addition, I thank the Aspen Center for Physics and the Simons workshop in Mathematics and Physics, Stony Brook 2009, for hospitality during the initial stages of this work. My research is supported by DOE grant FG02-84-ER40153.

\end{document}